# Performance Comparison Between VoLTE and non-VoLTE Voice Calls During Mobility in Commercial Deployment: A Drive Test-Based Analysis

Rashed Hasan Ratul
*Department of Electrical and Electronic Engineering*
*Islamic University of Technology (IUT)*
Dhaka, Bangladesh
rashedhasan@iut-dhaka.edu

Muhammad Iqbal
*Department of Communications Engineering*
*National Chung Cheng University (CCU)*
Chiayi, Taiwan
iqbalmarjan@alum.ccu.edu.tw

Jen-Yi Pan
*Department of Communications Engineering*
*National Chung Cheng University (CCU)*
Chiayi, Taiwan
jypan@ccu.edu.tw

Mohammad Mahadi Al Deen
*Department of Electrical and Electronic Engineering*
*Islamic University of Technology (IUT)*
Dhaka, Bangladesh
mahadialdeen@iut-dhaka.edu

Mohammad Tawhid Kawser
*Department of Electrical and Electronic Engineering*
*Islamic University of Technology (IUT)*
Dhaka, Bangladesh
kawser@iut-dhaka.edu

Mohammad Masum Billah
*Department of Electrical and Electronic Engineering*
*Islamic University of Technology (IUT)*
Dhaka, Bangladesh
masum.billah@iut-dhaka.edu

*Abstract*— The optimization of network performance is vital for the delivery of services using standard cellular technologies for mobile communications. Call setup delay and User Equipment (UE) battery savings significantly influence network performance. Improving these factors is vital for ensuring optimal service delivery. In comparison to traditional circuit-switched voice calls, VoLTE (Voice over LTE) technology offers faster call setup durations and better battery-saving performance. To validate these claims, a drive test was carried out using the XCAL drive test tool to collect real-time network parameter details in VoLTE and non-VoLTE voice calls. The findings highlight the analysis of real-time network characteristics, such as the call setup delay calculation, battery-saving performance, and DRX mechanism. The study contributes to the understanding of network optimization strategies and provides insights for enhancing the quality of service (QoS) in mobile communication networks. Examining VoLTE and non-VoLTE operations, this research highlights the substantial energy savings obtained by VoLTE. Specifically, VoLTE saves approximately 60.76% of energy before the Service Request and approximately 38.97% of energy after the Service Request. Moreover, VoLTE to VoLTE calls have a 72.6% faster call setup delay than non-VoLTE-based LTE to LTE calls, because of fewer signaling messages required. Furthermore, as compared to non-VoLTE to non-VoLTE calls, VoLTE to non-VoLTE calls offer an 18.6% faster call setup delay. These results showcase the performance advantages of VoLTE and reinforce its potential for offering better services in wireless communication networks.

## I. INTRODUCTION

In the rapidly evolving mobile communication sector, customers usually demand uninterrupted, efficient, and consistent service, especially with regard to call setup duration and battery savings. The need for constant and faster connectivity to operator services, including the internet, while moving has become increasingly important. The limitations of traditional circuit-switched phone calls can significantly impact the user experience in terms of call setup time and voice quality [1]. Moreover, during cell switching, the call setup delay often increases, leading to dissatisfied consumers. Therefore, the implementation of VoLTE service, which allows for faster call setup time and better utilization of battery savings, has become a crucial aspect of user demand [2]. The purpose of this research is to conduct a practical assessment of the network's performance in VoLTE enabled urban areas. The study aims to analyze the call setup delay performance and the coverage of the network using a standard field experiment using the XCAL drive test tool. By providing performance data, this research aims to assist network service providers, planners, and designers in making informed decisions regarding the integration, enhancement, and deployment of VoLTE service in their network infrastructure. With the widespread adoption of LTE by mobile network operators, VoLTE has emerged as the most favored technical standard for delivering faster and high-quality voice services over LTE networks [3]. By implementing VoLTE technology, mobile network operators no longer need to dedicate separate circuit-switched (CS) services for voice and data, thereby simplifying their network infrastructure and reducing operational costs [4]. This integration of voice and data services onto a single LTE network streamlines network management and delivers a more efficient, cost-effective, and seamless user experience.

Mobile communication networks using standardized cellular technology must optimize their network performance to ensure optimal service delivery. Call setup time and DRX mechanism are two critical performance metrics that must be taken into consideration. Real-time data that captures a variety of network characteristics affecting performance is required to confirm the benefits of VoLTE technology. While considerable research has been done on this subject, there is insufficiently clear and thorough documentation on these aspects. The primary objective of this study is to present a test strategy for analyzing the performance of call setup delay and overall relative battery saving ability in VoLTE compared to non-VoLTE operation. This article delves into the critical issue of the time taken to initiate a voice call. The aim is to shed light on the potential challenges and opportunities for improvement in network performance by conducting a comprehensive analysis of this fallback time. Through the implementation of VoLTE technology, it has been discovered that the additional time taken to fall back to 3G from LTE service can be greatly minimized, resulting in faster call initiation and a more satisfactory user experience [5].

This research article also focuses on testing new features of VoLTE service for utilizing better power-saving ability and verifying end-to-end system functionality. It addresses the challenges that test engineers may encounter during power-saving performance testing and call setup time, proposing an assessment strategy for field environment testing. Precise testing of the battery saving mechanism and call setup delays are necessary for valid results and further analysis. This research advances network optimization knowledge and recommends methods for enhancing mobile communication network quality of service (QoS). With a focus on the effects of signaling messages and the DRX cycle, the research seeks to figure out the factors that contribute to the shorter call setup delay as well as better utilization of the power-saving mechanism in VoLTE compared to traditional voice services.

## II. LITERATURE REVIEW

Krasniqi et al. examined VoLTE performance in real time, including voice quality, call setup time, and call reliability. In order to acquire real-time network parameters, the authors performed a drive test [6]. Ayman et al. investigated and analyzed Circuit Switched Fallback (CSFB) and VoLTE systems' call setup delay, handover protocols, and KPIs. The article also discusses Single Radio Voice Call Continuity (SRVCC) and its evolution. The performance of CSFB and VoLTE in terms of call setup delay and in-call mobility performance is also assessed. Finally, the article provided strategies for minimizing call setup duration and enhancing eSRVCC handover success [7]. However, there is no validated analysis of drive tests for assessing real-time data in the article. Ayman et al. also investigated VoLTE in commercial 3GPP Release-10 LTE networks. The study also provided deployment guidelines. The assessment took place on active commercial LTE networks under typical conditions and in scenarios involving mobility at an average speed of 80 km/h. The study concluded that VoLTE voice calls with a speech rate of 12.65 kbps provided improved voice quality to OTT and CS voice calls [8]. Aleksander et al. undertook a study to assess the performance of VoLTE for railway-specific voice communication, with an emphasis on one-to-one operational calls and Railway Emergency Calls (REC). The performance of VoLTE was tested against rail industry standards using simulation scenarios. The findings proved that the call setup process for VoLTE was quicker than what was required by railways. According to the findings, VoLTE adequately addresses railway communication requirements. However, the paper's generalization to other railway systems or regions may be limited, and the simulations may not have reflected real-world operational conditions [9].

Bautista et al. examined CSFB's performance in optimized LTE Rel-8 networks and M2M scenarios on live commercial LTE networks. The authors compared mobile-originated (MO) and mobile-terminated (MT) call delays to landline calls. Their studies showed that MO UE observed delay was twice as long as landlines, whereas MT UE perceived delay was equivalent to landline MT calls. The report also included CSFB call setup failure analysis, including network optimization, NAS message handling, and Release 9 CSFB features. It should be noted, however, that the study is limited to CSFB and may not be applicable to other communication techniques [10]. Pastrav et al. examined Voice over Internet Protocol (VoIP) QoS and QoE in an LTE cell deployed in a 3D-modeled Romanian metropolitan area. The simulation findings assist in planning an efficient LTE network based on area parameters, user density, and VoIP performance. However, the study's simulation-based evaluation may not fully reflect real-world applications and it focused only on VoIP performance without considering other applications or services [11]. Abichandani et al. examined coverage at pedestrian and vehicular speeds for various voice over LTE adaptive multi rate wideband codec mode-sets. The study included laboratory testing in an RF screen room with attenuated RF signals and outdoor testing located in rural Wisconsin on a level drive route at 90 km/h. The researchers tested numerous mode-sets using the XCAL-MPM4 box to record data from four mobile devices simultaneously. The goal was to assess the voice quality and coverage of various mode-sets in LTE networks. However, the study's restricted scope of testing in a controlled lab environment and a specific geographic region may not fully reflect practical network conditions [12]. The objective of Vehanen et al.'s thesis is to enhance the performance testing of Inter Radio Access Technology (I-RAT) handovers from LTE to 3G networks by developing a comprehensive test plan. The authors thoroughly explore the issues test engineers may face during I-RAT handover testing and offer a field environment test plan. The drive test was carried out with the support of XCAL [13]. Gadze et al. compared WiMAX network performance using simulation and field data. The study used GPS, Dongle XCAL-X, a laptop with XCAL-X software, and a van to assess distance and coverage in Accra's urban areas. The researchers employed varying downlink/uplink ratios to balance throughput during field experiments across the University of Ghana Campus to simulate urban outdoor wireless network challenges. The simulation and measurement results shed light on the theoretical and simulated performance of the MIMO configuration based on the network simulation parameters used at the test site [14].

One of the major challenges in studying and improving technological systems is the limited scope of research. Most studies tend to focus on a specific technology or scenario, which may not be directly applicable to other systems or regions. Another common issue in research is the lack of real-time evaluation, where some studies do not include real-time analysis of network parameters and instead rely on simulation scenarios or laboratory testing. This approach may not always reflect the real-world complexities of the system, and thus limit the applicability of the findings. Researchers must adopt comprehensive, dynamic approaches that encompass diverse scenarios and include real-time monitoring and analysis to overcome these limitations. The significance of our research lies in addressing the limitations of previous studies that have mainly focused on specific technologies or scenarios, with limited applicability to other systems or regions. Our research aims to bridge this gap by analyzing real-time signaling messages transmitted between UE and eNodeB, enabling us to accurately calculate the call setup delay and analyzing the power saving mechanism for VoLTE. This is a novel approach, as no previous research has explored the actual reason for the faster call setup time as well as the factor behind better power saving mechanisms together in VoLTE with proper validation and justification of real-time signaling messages.

The quantitative analysis of this research is critical in providing high-level accurate measurements of call setup delay analysis and the reason behind enhanced power saving. Additionally, we have conducted two-way voice calls simultaneously in different real-life scenarios, providing insights into the variability of signaling messages that have not been explored in current accessible research. Overall, our research aims to enhance the technical understanding of call setup delay and power-saving techniques in VoLTE and contribute to the development of more efficient and reliable communication systems.

## III. METHODOLOGY

The evaluation of VoLTE service from an end-user experience perspective involves conducting mobility tests at driving speeds while collecting real-time network data. In this study, measurements were obtained through measurement operations conducted by a reputable mobile operator in Bangladesh, where VoLTE technology is extensively deployed. The measurements were performed in Gulshan, a large urban region encompassing connecting roads between sectors and adjacent smaller communities. The aim was to assess the technical effectiveness of the network, specifically focusing on end users equipped with VoLTE-capable smartphones. By gathering numerical measurements and analyzing their statistical distributions, valuable insights were obtained regarding the performance and quality of the VoLTE service in real-world mobility scenarios.

### A. Drive Test Plan

The tests are performed utilizing two separate mobile-to-mobile systems in order to call one another. The focus of this activity is to monitor the network messages and signal information that occurs during the call setup process. Additionally, each of these analyses is performed for VoLTE to VoLTE calls, VoLTE to regular LTE calls, and LTE to LTE calls, enabling a thorough comparison of call setup performance between various technologies. The strategy employed in this study comprises measuring mobility in a real-world network setting in order to get empirical information on VoLTE service performance.

The drive test plan was meticulously designed to assess the performance of VoLTE service in a real-world urban setting. A specially equipped microbus was utilized for conducting the tests, ensuring mobility and coverage across Gulshan 1 and 2, bustling metropolitan areas in Bangladesh. The tests spanned an entire day, from 10.30 am to 4.30 pm, capturing network data under diverse environmental conditions and traffic scenarios. To capture and analyze network parameters, signaling messages, and technical insights, two mobile phones, namely Samsung Galaxy S10 and Samsung Galaxy S10+, were connected to a laptop equipped with the XCAL drive test software. This licensed software, widely recognized for its reliability in assessing mobile network performance, facilitated real-time monitoring and data collection during phone calls [15].

Through the XCAL drive test software, an extensive range of technical aspects was examined, including call setup delay, signal strength, handover performance, and voice quality metrics. This comprehensive analysis allowed for the evaluation of key performance indicators (KPIs) specific to VoLTE and non-VoLTE operations, such as the time required for call setup, and the efficiency of battery-saving techniques. By conducting tests in a dynamic urban environment and leveraging advanced measurement capabilities, this study provided valuable insights into the technical capabilities and performance of VoLTE. Both mobile phones were equipped with VoLTE-supported SIM cards from the same operator, which allowed for testing VoLTE service performance in a real-world network environment.

### B. Dynamic Mobility Scenario Evaluation

Each test conducted as part of this study involved a comprehensive evaluation of VoLTE service performance in dynamic mobility scenarios. The tests were carried out for approximately 30 minutes, encompassing a single drive loop with the microbus traveling at an average speed of 45 km/h. This speed was chosen to simulate typical vehicular speeds, ensuring that the tests captured the real-world performance of VoLTE service during on-the-go scenarios. By conducting the tests while the vehicle was in motion, the study aimed to assess the impact of mobility on VoLTE performance. To maintain consistency and comparability across measurement scenarios, a predetermined routing path was followed for all tests. This routing path, illustrated in Fig. 1, provided a standardized framework for assessing the performance of VoLTE service under varying network conditions and common geographic locations in most of the urban areas with VoLTE enabled operations.

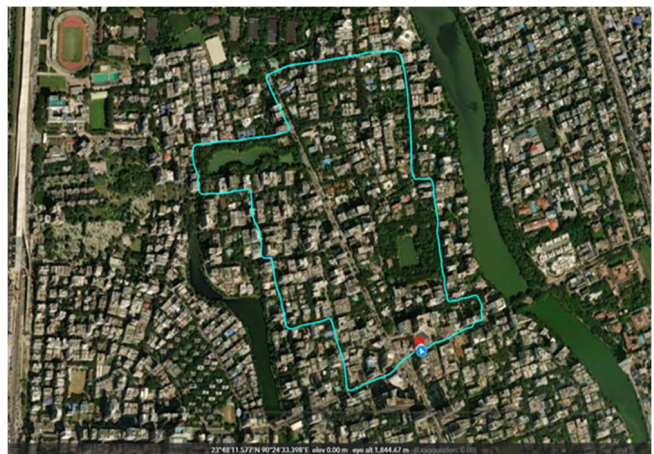

Fig. 1. Routing path for the drive test.

### C. XCAL Software Tool for Field Test

The XCAL drive test software used in the study provided detailed signaling message notifications at the millisecond level, making it easier to accurately calculate call setup delays for each cellular communication technology [16]. The precise timing information captured by the software allowed for precise measurement of call setup delays for VoLTE and general circuit-switching non-VoLTE oriented operations, enabling a comprehensive analysis of the network behavior and performance during call setup. This level of granularity in the signaling message notifications provided by XCAL software facilitated the accurate calculation of call setup delays for each case, ensuring reliable results for the drive test. Overall, the drive test plan aimed to gather empirical data on the performance of VoLTE service in a real-world network environment, using a van equipped with mobile phones and specialized drive test software. The collected data and observations from different scenarios would provide valuable insights into the technical performance of VoLTE service, contributing to a comprehensive analysis of VoLTE service performance. The XCAL setup is shown in Fig. 2.

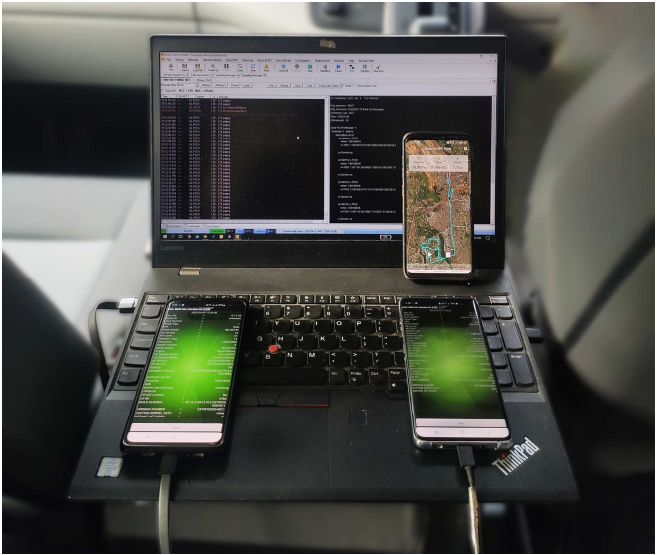

Fig. 2. XCAL drive test tool setup with cell phones.

## IV. OVERALL ANALYSIS

This section presents a comprehensive analysis of key variables that play a crucial role in the performance of VoLTE service. By examining these variables in detail, valuable insights can be gained regarding the optimization of VoLTE network implementation and deployment.

### A. Call Setup Delay

Call setup delay refers to the time taken for a voice call to be established from the initiation of the call to the point where the call is connected and the parties can communicate. Call setup delay is one of the key performance metrics that is measured and analyzed to assess the performance of VoLTE service [17]. Different scenarios are observed for calls made from one mobile to another mobile and vice versa, and for different types of calls, including VoLTE to VoLTE calls, VoLTE to traditional LTE calls, and LTE to LTE calls. The objective is to capture different call setup scenarios and analyze the network behavior during call setup. Fig. 3 illustrates the sample signaling messages after initiating a phone call as well as the call setup delay for VoLTE to VoLTE service.

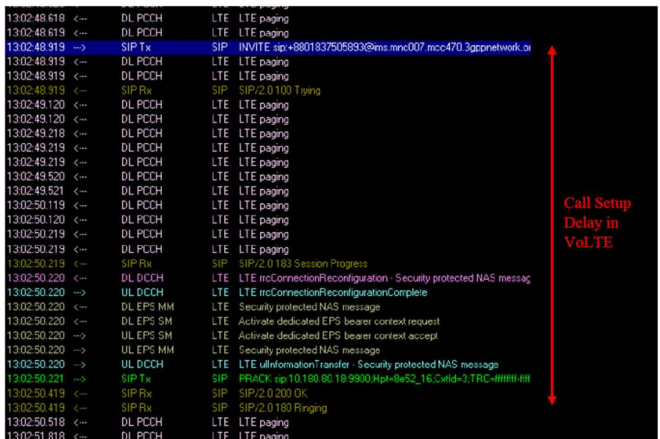

Fig. 3. Observed signaling messages for VoLTE to VoLTE call.

These statistical metrics help in understanding the typical call setup performance of the VoLTE network, including the time taken for call setup, the variability in call setup delay, and any potential issues or optimizations needed. The call setup delay data is also compared with traditional CSFB technologies to understand the differences in call setup performance between VoLTE and these technologies.

Cellular communication technologies have gone through numerous stages to offer advanced features [18]. Hence, by understanding the call setup delay behavior, network operators and service providers can identify any performance issues, optimize network configurations, and ensure a satisfactory and advanced user experience for VoLTE service users [19]. Table 1 shows the simplified sample signaling messages after initiating a phone call for non-VoLTE operation. Call setup delay for non-VoLTE service has been calculated using the time gap between the Extended service request message and the ALTERING message.

TABLE I. SIMPLIFIED SIGNALING MESSAGES FOR NON-VOLTE TO NON-VOLTE VOICE CALLS[1]

| Time [2] | UE-NET | ID | Message |
|---|---|---|---|
| 13:31:30.743 | <--- | LTE | LTE paging |
| **13:31:30.843** | **<---** | **LTE** | **Extended service request** |
| 13:31:30.843 | <--- | LTE | Security protected NAS message |
| 13:31:31.469 | <--- | NAS | CM SERVICE REQUEST |
| 13:31:31.567 | <--- | 3G | rrcConnectionRequest - originatingConversationalCall |
| 13:31:31.567 | <--- | 3G | pagingType1 - terminatingHighPrioritySignalling[3] |
| 13:31:32.067 | <--- | 3G | measurementControl[4] |
| 13:31:32.170 | <--- | NAS | Routing area update request |
| 13:31:32.268 | <--- | 3G | measurementReport |
| 13:31:32.268 | ---> | 3G | initialDirectTransfer - Routing area update request |
| 13:31:32.269 | ---> | NAS | SETUP |
| 13:31:32.269 | ---> | 3G | uplinkDirectTransfer - SETUP |
| 13:31:32.469 | <--- | 3G | measurementReport |
| 13:31:32.668 | ---> | 3G | downlinkDirectTransfer - CALL PROCEEDING |
| 13:31:32.668 | <--- | NAS | CALL PROCEEDING |
| 13:31:33.368 | <--- | 3G | measurementReport - e6b[5] |
| 13:31:36.469 | ---> | 3G | radioBearerReconfigurationComplete |
| 13:31:36.470 | ---> | 3G | downlinkDirectTransfer - ALERTING |
| **13:31:36.470** | **<---** | **NAS** | **ALERTING** |
| 13:31:37.671 | <--- | 3G | measurementReport - e2f |

Four sets of signaling messages play a crucial role in VoLTE-VoLTE call setup which has been shown in Table 2. The timely exchange of these messages determines the overall call setup delay, which is a critical performance metric in the VoLTE service. Therefore, to ensure optimal call setup time, it is essential to ensure that these four sets of signaling messages are delivered in a timely and efficient manner.

TABLE II. FOUR SETS OF COMMON SIGNALING MESSAGES IN VOLTE ENABLED VOICE CALLS.

| UE-NET | Network ID | Signaling Message |
|---|---|---|
| <--- | LTE | Security protected NAS message |
| <--- | LTE | Activate EPS bearer context request |
| ---> | LTE | Activate EPS bearer context accept |
| ---> | LTE | Security protected NAS message |

---

[1] In actual case, the signaling messages are much more in number than shown here for simplified representation
[2] Time is given here as *hh:mm:ss:ms* format
[3] Approximately 18 pagingType1 messages transferred
[4] Approximately 6 measurementControl messages transferred
[5] Multiple measurementReports and radioBearer messages transferred

## B. DRX Cycle and Power Saving Mechanism Analysis

The variation in the DRX cycle during VoLTE call setup signifies the adaptive power-saving operation of the system. When the time difference between signaling messages is high, the UE can switch to a longer DRX cycle to reduce the number of times it needs to wake up and receive messages from the eNodeB, thus saving power. Conversely, when the time difference between signaling messages is low, the UE can switch to a shorter DRX cycle to ensure the timely receipt of messages and prevent call setup delays [20]. This adaptive mechanism allows VoLTE to efficiently manage battery usage during call setup and other communication operations. Fig. 4 shows the variation of the DRX cycles between VoLTE and non-VoLTE services.

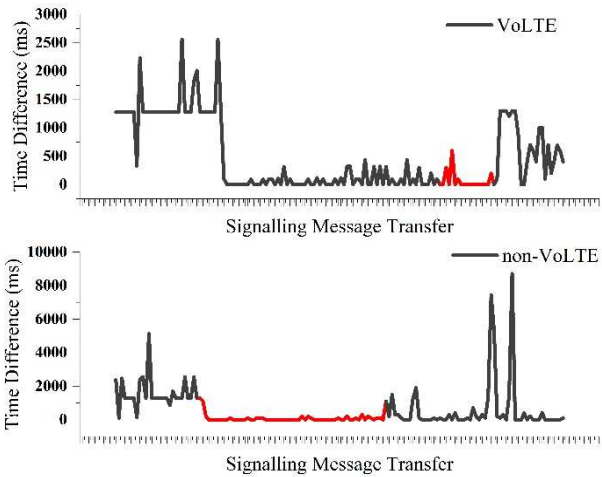

Fig. 4. Variation of DRX cycle during Call Setup Delay (in Red).

During a VoLTE call setup, there is a significant variation in the time difference between the transmission and reception of signaling messages between UE and eNodeB. This leads to a sharp rise and drop in the DRX cycle, which is an adaptive mechanism used by VoLTE to save power [21]. The sharp variation of the DRX cycle in VoLTE enabled voice calls is shown in Fig. 5.

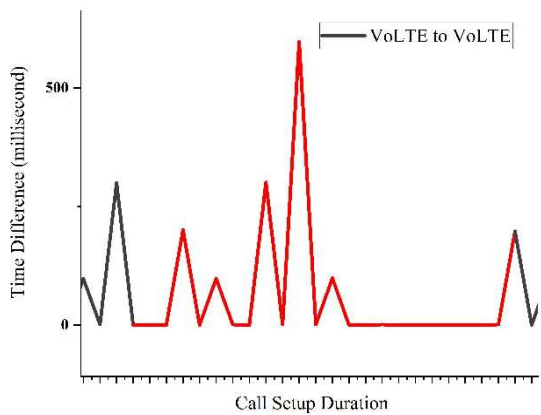

Fig. 5. Time variation of signaling messages during call setup in VoLTE.

In contrast, non-VoLTE operation has a nearly steady DRX cycle and fixed time difference between signaling messages as shown in Fig. 4, resulting in continuous signaling messages and reduced probability of battery savings. The adaptive DRX cycle mechanism in VoLTE call setup enables efficient battery usage and improves the overall performance of the system [22]. Furthermore, our research also revealed that the power-saving performance of VoLTE is significantly better than traditional voice services. This is due to the adaptive DRX cycle mechanism employed by VoLTE, which adjusts the cycle duration based on the quality of the radio link and the overall situation of user activity before and after the initiation of any Service Request by the UE. This results in reduced power consumption during periods of inactivity, without compromising the call quality or setup time.

## C. Paging Messages for VoLTE and non-VoLTE

Paging messages are essential in communication networks for locating and notifying mobile devices of incoming calls or messages. They contain important information and are broadcasted to reach all devices within a coverage area [23]. Mobile devices respond to paging messages to confirm their presence and readiness for connection. Efficient handling of paging messages improves network performance and conserves battery life [24]. Advancements in LTE and 5G networks have introduced enhancements to paging mechanisms, optimizing efficiency and user experience. The time interval of paging messages in a communication network is closely related to the DRX mechanism and battery saving [25]. DRX is a power-saving technique used in cellular networks where the mobile device periodically enters sleep mode to conserve battery. During sleep mode, the device disables certain functionalities, including continuous monitoring of the paging channel. Instead, it wakes up periodically to check for paging messages within predefined intervals [26]. The time interval of paging messages directly affects the power consumption of mobile devices [27]. If the paging interval is short, the device needs to wake up more frequently to check for incoming calls or messages. This increased wake-up frequency results in higher power consumption and potentially reduces battery life.

On the other hand, if the paging interval is long, the device can remain in sleep mode for extended periods, reducing power consumption and conserving battery life [28]. However, a longer paging interval may lead to delays in receiving incoming calls or messages, as the device checks for them less frequently. Thus, there exists a trade-off between battery saving and timely message reception. Adjusting the paging interval requires finding the right balance between power efficiency and the need for timely communication. Networks strive to optimize this balance by considering factors such as user preferences, network congestion, and the urgency of incoming communications. The research findings provide valuable insights into the characteristics of paging message time intervals in VoLTE and non-VoLTE operations. This study reveals that in VoLTE, the paging message time interval initially exhibits a higher duration prior to receiving the Service Request message at the eNodeB. However, upon receiving the Service Request from the UE, there is a significant reduction in the signaling time interval. This reduction in the paging message time gap after the Service Request message reflects an adaptive process specifically designed for VoLTE, aiming to minimize call setup delays.

The adaptive nature of the paging message time interval in VoLTE showcases an important optimization mechanism. It effectively reduces the occurrence of unnecessary signaling message transfers, thereby streamlining call setup procedures and improving overall efficiency. In addition, this adaptive strategy aids in reducing battery consumption. The observation of high time gaps initially suggests the presence of excessive signaling message transfers, which are effectively mitigated through the implementation of an adaptive paging message time interval in VoLTE. These techniques allow for longer sleep durations and more flexible paging strategies, reducing unnecessary wake-ups and extending battery life without significantly impacting the responsiveness of the device. Fig. 6 showcases the paging message time intervals in the VoLTE case. It visually demonstrates the initial higher interval before receiving the Service Request, followed by a significant decrease in the signaling time interval. This adaptive feature in VoLTE reduces call setup delays and enhances energy efficiency.

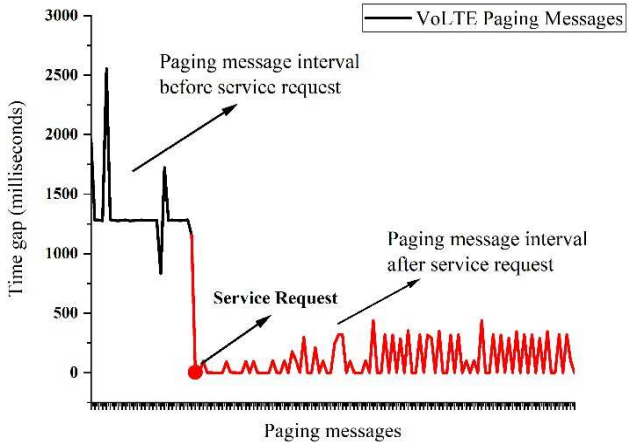

Fig. 6. Paging message interval analysis in VoLTE.

Conversely, the analysis emphasizes that non-VoLTE operations lack this advanced feature of an adaptive paging message time interval both before and after the Service Request. As a result, the battery-saving mechanism in non-VoLTE operations is comparatively less user-friendly, potentially leading to increased power consumption and reduced battery efficiency. Fig. 7 displays the paging message time intervals in the non-VoLTE case. Unlike VoLTE, the graph shows a static interval without an adaptive mechanism. This absence of adaptivity in non-VoLTE systems may result in less optimized call setup delays and potential inefficiencies in battery usage. These research findings underscore the significant advantages associated with VoLTE's adaptive paging message time interval. By optimizing call setup delays and improving energy efficiency, VoLTE offers an enhanced user experience compared to non-VoLTE systems. The adaptive approach ensures a more streamlined and efficient signaling process, benefiting both call setup performance and battery consumption at the UE end. The percentage of energy savings can be calculated using the following formula [29]:

$$\text{Percentage energy savings} = \frac{ME_{sleep} + NE_{awake}}{(M+N)E_{awake}} \quad (1)$$

In this case, the system considers a scenario where the UE operates in two distinct modes: the DRX mode and the normal operation mode. The DRX mode consists of $M$ frames, during which the UE is in a power-saving state, while the normal operation mode comprises $N$ frames, where the UE operates in its regular mode without power-saving mechanisms.

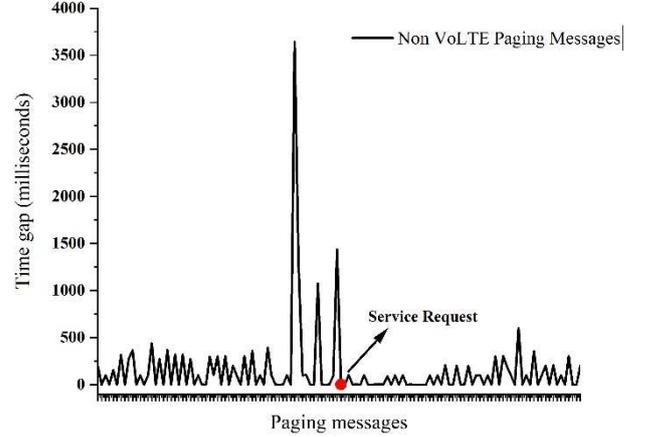

Fig. 7. Paging message interval analysis in non-VoLTE.

Our proposed formula for calculating energy savings based on the time interval of signaling messages is as follows:

$$\text{Energy savings} = \frac{MT_{interval}*E_{per\ unit} + NT_{awake}*E_{per\ unit}}{(M+N)T_{awake}*E_{per\ unit}} \quad (2)$$

Here, the equation under consideration involves the variables $M$ and $N$, which symbolize the total number of signaling blocks before and after the initiation of the Service Request message, respectively. $T_{interval}$ and $T_{awake}$ are employed to represent the average time gap between two corresponding signaling messages before and after the initiation of the Service Request message. Lastly, $E_{per\ unit}$ denotes the average energy consumed by the battery for the transmission of each signaling block.

## V. RESULTS

The call setup delay time for VoLTE-VoLTE calls is almost 72.6% faster compared to LTE-LTE calls. This difference is primarily due to the reduced number of signaling messages involved in VoLTE-VoLTE calls. Moreover, optimum delivery of the message blocks associated with EPS bearer context request message during VoLTE service enables faster call setup in VoLTE-VoLTE calls. On the other hand, traditional LTE-LTE calls require additional message blocks while switching back to 3G, resulting in longer call setup times. Even when making calls from a VoLTE user to a non-VoLTE user, the average call setup delay is significantly faster than traditional circuit-switched voice calls. Our findings reveal that the average call setup delay for VoLTE to non-VoLTE calls is 18.6% faster than the average call setup delay for non-VoLTE to non-VoLTE calls. In terms of energy efficiency, VoLTE exhibits a significant energy saving of approximately 60.76% compared to non-VoLTE before the Service Request initiation. Similarly, after the Service Request, VoLTE showcases a notable energy saving of approximately 38.97% compared to non-VoLTE, highlighting its efficiency in optimizing energy consumption during signaling and communication processes.

## VI. Conclusion

VoLTE, a highly optimized and advanced cellular technology, offers numerous advantages over LTE and other traditional cellular networks. It excels in terms of battery-saving techniques, making it a more efficient and user-friendly choice. VoLTE also outperforms traditional cellular networks when considering call setup delays. With its advanced battery-saving techniques and adaptive features, VoLTE ensures enhanced battery performance, resulting in extended device usage and improved user experience. In conclusion, VoLTE surpasses LTE and traditional cellular technologies with its optimized energy consumption, faster call setup technique, and advanced battery-saving procedures. These advantages represent VoLTE as a more sustainable, efficient, and user-friendly choice for modern communication networks.

## Acknowledgment

The drive tests were performed using the licensed XCAL software with sponsored logistical support from IUT, Bangladesh. This work is also supported in part by Industrial Technology Research Institute (ITRI), Taiwan.